# Cross-Composition: A New Technique for Kernelization Lower Bounds*

## Hans L. Bodlaender[1], Bart M. P. Jansen[1], and Stefan Kratsch[1]

1   Department of Information and Computing Sciences, Utrecht University
    P.O. Box 80.089, 3508 TB, Utrecht, The Netherlands
    {hansb,bart,kratsch}@cs.uu.nl

──── Abstract ────

We introduce a new technique for proving kernelization lower bounds, called *cross-composition*. A classical problem $L$ cross-composes into a parameterized problem $Q$ if an instance of $Q$ with polynomially bounded parameter value can express the logical OR of a sequence of instances of $L$. Building on work by Bodlaender et al. (ICALP 2008) and using a result by Fortnow and Santhanam (STOC 2008) we show that if an NP-hard problem cross-composes into a parameterized problem $Q$ then $Q$ does not admit a polynomial kernel unless the polynomial hierarchy collapses.

Our technique generalizes and strengthens the recent techniques of using OR-composition algorithms and of transferring the lower bounds via polynomial parameter transformations. We show its applicability by proving kernelization lower bounds for a number of important graphs problems with structural (non-standard) parameterizations, e.g., CHROMATIC NUMBER, CLIQUE, and WEIGHTED FEEDBACK VERTEX SET do not admit polynomial kernels with respect to the vertex cover number of the input graphs unless the polynomial hierarchy collapses, contrasting the fact that these problems are trivially fixed-parameter tractable for this parameter. We have similar lower bounds for FEEDBACK VERTEX SET.

## 1   Introduction

Preprocessing and data reduction are important and widely applied concepts for speeding up polynomial-time algorithms or for making computation feasible at all in the case of hard problems that are not believed to have efficient algorithms. Kernelization is a way of formalizing data reduction, which allows for a formal analysis of the (im)possibility of data reduction and preprocessing. It originated as a technique to obtain fixed-parameter tractable algorithms for hard (parameterized) problems, and has evolved into its own topic of research (see [19, 2] for recent surveys). A *parameterized problem* [14, 16] is a language $Q \subseteq \Sigma^* \times \mathbb{N}$, the second component is called the *parameter*. A *kernelization* algorithm (*kernel*) transforms an instance $(x, k)$ in polynomial time into an equivalent instance $(x', k')$ such that $|x'|, k' \leq f(k)$ for some computable function $f$, which is the *size* of the kernel.

From a practical perspective we are particularly interested in cases where $f \in k^{O(1)}$, so-called *polynomial kernels*. Success stories of kernelization include the $O(k^2)$ kernel for $k$-VERTEX COVER containing at most $2k$ vertices [11] and the meta-theorems for kernelization

────────────

* This work was supported by the Netherlands Organisation for Scientific Research (NWO), project "KERNELS: Combinatorial Analysis of Data Reduction".

of problems on planar graphs [4], among many others (cf. also [23]). Although researchers have looked for polynomial kernels for elusive problems such as $k$-Path for many years, it was only recently that techniques were introduced which make it possible to prove (under some complexity-theoretic assumption) that a parameterized problem in FPT does not admit a polynomial kernel. Bodlaender et al. [3] introduced the concept of a or-*composition* algorithm as a tool to give super-polynomial lower bounds on kernel sizes. Consider some set $S$, and let $\textsc{or}(S)$ denote the set such that for any sequence $x^* := (x_1, \ldots, x_t)$ of instances of $S$ we have $x^* \in \textsc{or}(S) \Leftrightarrow \bigvee_{i=1}^{\ell} x_i \in S$; then we could say that the language $\textsc{or}(S)$ expresses the or of instances of $S$. The approach taken in the original paper by Bodlaender et al. [3] uses a theorem by Fortnow and Santhanam [17] to show that if there is a polynomial-time or-composition algorithm that maps any sequence of instances $(x_1, k), (x_2, k), \ldots, (x_t, k)$ of some parameterized problem $Q$ which all share the same parameter value to an instance $(x^*, k^*)$ of $Q$ which acts as the or of the inputs and $k^* \in k^{O(1)}$, then $Q$ does not admit a polynomial kernel unless $\text{NP} \subseteq \text{coNP/poly}$. This machinery made it possible to prove e.g. that $k$-Path and the Clique problem parameterized by the treewidth of the graph do not admit polynomial kernels unless $\text{NP} \subseteq \text{coNP/poly}^1$. The latter is deemed unlikely since it is known to imply a collapse of the polynomial hierarchy to its third level [25] (and further [8]).

It did not take long before the techniques of Bodlaender et al. were combined with the notion of a *polynomial parameter transformation* to also prove lower bounds for problems for which no direct or-composition algorithm could be found. This idea was used implicitly by Fernau et al. [15] to show that $k$-Leaf Out-Branching does not admit a polynomial kernel, and was formalized in a paper by Bodlaender et al. [6]: they showed that if there is a polynomial-time transformation from $P$ to $Q$ which incurs only a polynomial blow-up in the parameter size, then if $P$ does not admit a polynomial kernel then $Q$ does not admit one either. These polynomial parameter transformations were used extensively by Dom et al. [13] who proved kernelization lower bounds for a multitude of important parameterized problems such as Small Universe Hitting Set and Small Universe Set Cover. Dell and van Melkebeek [12] were able to extend the techniques of Fortnow and Santhanam to prove, e.g., that Vertex Cover does not admit a kernel of size $O(k^{2-\epsilon})$ for any $\epsilon > 0$.

**Our results.** We introduce a new technique to prove kernelization lower-bounds, which we call *cross-composition*. This technique generalizes and strengthens the earlier methods of or-composition [3] and polynomial-parameter transformations [6], and puts the two existing methods of showing kernelization lower bounds in a common perspective. Whereas the existing notion of or-composition works by composing multiple instances of a *parameterized* problem $Q$ into a single instance of $Q$ with a bounded parameter value, for our new technique it is sufficient to compose the or of any *classical* NP-hard problem into an instance of the parameterized problem $Q$ for which we want to prove a lower-bound. The term *cross* in the name stems from this fact: the source- and target problem of the composition need no longer be the same. Since the input to a cross-composition algorithm is a list of *classical* instances instead of parameterized instances, the inputs do not have a parameter in which the output parameter of the composition must be bounded; instead we require that the size of the output parameter is polynomially bounded in the size of the largest input instance. In addition we show that the output parameter may depend polynomially on the logarithm of the number of input instances, which often simplifies the constructions and proofs. We also introduce the concept of a *polynomial equivalence relation* to remove the need for padding

---

| Problem name | Parameter | Kernel size | |
|---|---|---|---|
| CLIQUE | vertex cover | not polynomial | [Section 4.1] |
| CHROMATIC NUMBER | vertex cover | not polynomial | [Section 4.2] |
| FEEDBACK VERTEX SET | dist. from cluster | not polynomial | [Section 4.3] |
| FEEDBACK VERTEX SET | dist. from co-cluster | not polynomial | [Section 4.3] |
| WEIGHTED FVS | vertex cover | not polynomial | [Section 4.3] |

**Table 1** An overview of the kernelization lower bounds obtained in this paper; all listed problems are fixed-parameter tractable with respect to this parameterization. Section 4 describes the parameterized problems in more detail.

arguments which were frequently required for OR-compositions.

To show the power of cross-composition we give kernelization lower bounds for *structural* parameterizations of several important graph problems. Since many combinatorial problems are easy on graphs of bounded treewidth [5], and since the treewidth of a graph is bounded by the vertex cover number, it is often thought that almost all problems become tractable when parameterized by the vertex cover number of the graph. We show that this is not the case for kernelization: CLIQUE, CHROMATIC NUMBER and WEIGHTED FEEDBACK VERTEX SET do not admit polynomial kernels parameterized by the vertex cover number of the graph. In the case of CLIQUE it was already known [3] that the problem does not admit a polynomial kernel parameterized by the treewidth of the graph; since the vertex cover number is at least as large as the treewidth we prove a stronger result. For the unweighted FEEDBACK VERTEX SET problem, which admits a polynomial kernel parameterized by the target size of the feedback set [24], we show that there is no polynomial kernel for the parameterization by deletion distance to cluster graphs or co-cluster graphs.

**Organization.** The paper is organized as follows. We first give some preliminary definitions. Section 3 gives the formal definition of cross-composition, and proves that cross-compositions allow us to give kernelization lower bounds. In Section 4 we apply the new technique to obtain kernelization lower bounds for various problems.

## 2 Preliminaries

In this work we only consider undirected, finite, simple graphs. Let $G$ be a graph and denote its vertex set by $V(G)$ and the edge set by $E(G)$. We use $\chi(G)$ to denote the chromatic number of $G$. If $V' \subseteq V(G)$ then $G[V']$ denotes the subgraph of $G$ induced by $V'$. A graph is a *cluster graph* if every connected component is a clique. A graph is a *co-cluster graph* if it is the edge-complement of a cluster graph. Throughout this work we use $\Sigma$ to denote a finite alphabet, but note that multiple occurrences of $\Sigma$ may refer to different alphabets. For positive integers $n$ we define $[n] := \{1, \dots, n\}$. The satisfiability problem for boolean formulae is referred to as SAT. For completeness we give the following core definitions of parameterized complexity [3, 14].

▶ **Definition 1.** A *parameterized problem* is a language $Q \subseteq \Sigma^* \times \mathbb{N}$, and is contained in the class (strongly uniform) FPT (for Fixed-Parameter Tractable) if there is an algorithm that decides whether $(x, k) \in Q$ in $f(k)|x|^{O(1)}$ time for some computable function $f$.

▶ **Definition 2.** A *kernelization* algorithm [19, 2], or in short, a *kernel* for a parameterized problem $Q \subseteq \Sigma^* \times \mathbb{N}$ is an algorithm that given $(x, k) \in \Sigma^* \times \mathbb{N}$ outputs in $p(|x| + k)$ time a pair $(x', k') \in \Sigma^* \times \mathbb{N}$ such that:

- $(x, k) \in Q \Leftrightarrow (x', k') \in Q$,
- $|x'|, k' \leq f(k)$,

where $f$ is a computable function, and $p$ a polynomial. Any function $f$ as above is referred to as the size of the kernel; if $f$ is a polynomial then we have a *polynomial kernel*.

## 3 Cross-Composition

### 3.1 The Definition

In this section we define the concept of cross-composition and give all the terminology needed to apply the technique.

▶ **Definition 3** (Polynomial equivalence relation). An equivalence relation $\mathcal{R}$ on $\Sigma^*$ is called a *polynomial equivalence relation* if the following two conditions hold:

1. There is an algorithm that given two strings $x, y \in \Sigma^*$ decides whether $x$ and $y$ belong to the same equivalence class in $(|x| + |y|)^{O(1)}$ time.
2. For any finite set $S \subseteq \Sigma^*$ the equivalence relation $\mathcal{R}$ partitions the elements of $S$ into at most $(\max_{x \in S} |x|)^{O(1)}$ classes.

▶ **Definition 4** (Cross-composition). Let $L \subseteq \Sigma^*$ be a set and let $Q \subseteq \Sigma^* \times \mathbb{N}$ be a parameterized problem. We say that $L$ *cross-composes* into $Q$ if there is a polynomial equivalence relation $\mathcal{R}$ and an algorithm which, given $t$ strings $x_1, x_2, \ldots, x_t$ belonging to the same equivalence class of $\mathcal{R}$, computes an instance $(x^*, k^*) \in \Sigma^* \times \mathbb{N}$ in time polynomial in $\sum_{i=1}^{t} |x_i|$ such that:

1. $(x^*, k^*) \in Q \Leftrightarrow x_i \in L$ for some $1 \leq i \leq t$,
2. $k^*$ is bounded by a polynomial in $\max_{i=1}^{t} |x_i| + \log t$.

### 3.2 How Cross-compositions Imply Lower Bounds

The purpose of this section is to prove that cross-compositions imply kernelization lower bounds. To give this proof we need some concepts from earlier work [3, 17, 12].

▶ **Definition 5** ([17]). A *weak distillation* of SAT into a set $L \subseteq \Sigma^*$ is an algorithm that:
- receives as input a sequence $(x_1, \ldots, x_t)$ of instances of SAT,
- uses time polynomial in $\sum_{i=1}^{t} |x_i|$,
- and outputs a string $y \in \Sigma^*$ with
  1. $y \in L \Leftrightarrow x_i \in $ SAT for some $1 \leq i \leq t$,
  2. $|y|$ is bounded by a polynomial in $\max_{i=1}^{t} |x_i|$.

▶ **Theorem 6** (Theorem 1.2 [17]). *If there is a weak distillation of* SAT *into any set* $L \subseteq \Sigma^*$ *then* $NP \subseteq coNP/poly$ *and the polynomial-time hierarchy collapses to the third level (PH = $\Sigma_3^p$).*

▶ **Definition 7** ([12]). The OR *of a language* $L \subseteq \Sigma^*$ is the set OR$(L)$ that consists of all tuples $(x_1, \ldots, x_t)$ for which there is an index $1 \leq i \leq t$ with $x_i \in L$.

▶ **Definition 8** ([3]). We associate an instance $(x, k)$ of a parameterized problem with the *unparameterized instance* formed by the string $x\#1^k$, where $\#$ denotes a new character that we add to the alphabet and 1 is an arbitrary letter in $\Sigma$. The *unparameterized version* of a parameterized problem $Q$ is the language $\tilde{Q} = \{x\#1^k \mid (x, k) \in Q\}$.

▶ **Theorem 9.** *Let $L \subseteq \Sigma^*$ be a set which is NP-hard under Karp reductions. If $L$ cross-composes into the parameterized problem $Q$ and $Q$ has a polynomial kernel then there is a weak distillation of* SAT *into* $\text{OR}(\tilde{Q})$ *and* NP $\subseteq$ coNP/poly.

**Proof.** The proof is by construction and generalizes the concepts of Bodlaender et al. [3]. Assuming the conditions in the statement of the theorem hold, we show how to build an algorithm which distills SAT into $\text{OR}(\tilde{Q})$. By the definition of cross-composition there is a polynomial equivalence relation $\mathcal{R}$ and an algorithm $C$ which composes $L$-instances belonging to the same class of $\mathcal{R}$ into a $Q$-instance.

The input to the distillation algorithm consists of a sequence $(x_1, \ldots, x_t)$ of instances of SAT, which we may assume are elements of $\Sigma^*$. Define $m := \max_{j=1}^{t} |x_j|$. If $t > (|\Sigma|+1)^m$ then there must be duplicate inputs, since the number of distinct inputs of length $m' \leq m$ is $|\Sigma|^{m'}$. By discarding duplicates we may therefore assume that $t \leq (|\Sigma| + 1)^m$, i.e., $\log t \in O(m)$. By the assumption that $L$ is NP-hard under Karp reductions, there is a polynomial-time reduction from SAT to $L$. We use this reduction to transform each SAT instance $x_i$ for $1 \leq i \leq t$ into an equivalent $L$-instance $y_i$. Since the transformation takes polynomial time, it cannot increase the size of an instance by more than a polynomial factor and therefore $|y_i|$ is polynomial in $m$ for all $i$.

The algorithm now pairwise compares instances using the polynomial-time equivalence test of $\mathcal{R}$ (whose existence is guaranteed by Definition 3) to partition the $L$-instances $(y_1, \ldots, y_t)$ into partite sets $Y_1, \ldots, Y_r$ such that all instances from the same partite set are equivalent under $\mathcal{R}$. The properties of a polynomial equivalence relation guarantee that $r$ is polynomial in $m$ and that this partitioning step takes polynomial time in the total input size.

We now use the cross-composition algorithm $C$ on each of the partite sets $Y_1, \ldots, Y_r$, which is possible since all instances from the same set are equivalent under $\mathcal{R}$. Let $(z_i, k_i)$ be the result of applying $C$ to a sequence containing the contents of the set $Y_i$, for $1 \leq i \leq r$. From the definition of cross-composition and using $\log t \in O(m)$ it follows that each $k_i$ is polynomial in $m$, and that the computation of these parameterized instances takes polynomial time in the total input size. From Definition 4 it follows that $(z_i, k_i)$ is a YES instance of $Q$ if and only if one of the instances in $Y_i$ is a YES instance of $L$, which in turn happens if and only if one of the inputs $x_i$ is a YES instance of SAT.

Let $K$ be a polynomial kernelization algorithm for $Q$, whose existence we assumed in the statement of the theorem. We apply $K$ to the instance $(z_i, k_i)$ to obtain an equivalent instance $(z'_i, k'_i)$ of $Q$ for each $1 \leq i \leq r$. Since $K$ is a polynomial kernelization we know that these transformations can be carried out in polynomial time and that $|z'_i|, k'_i \leq k_i^{O(1)}$. Since $k_i$ is polynomial in $m$ it follows that $|z'_i|$ and $k'_i$ are also polynomial in $m$ for $1 \leq i \leq r$.

As the next step we convert each parameterized instance $(z'_i, k'_i)$ to the unparameterized variant $\tilde{z}_i := z'_i \# 1^{k'_i}$. Since the values of the parameters are polynomial in $m$ this transformation takes polynomial time, and afterwards we find that $|\tilde{z}_i|$ is polynomial in $m$ for each $1 \leq i \leq r$.

The last stage of the algorithm simply combines all unparameterized variants into one tuple $x^* := (\tilde{z}_1, \tilde{z}_2, \ldots, \tilde{z}_r)$. Since the size of each component is polynomial in $m$, and since the number of components $r$ is polynomial in $m$, we have that $|x^*|$ is polynomial in $m$. The tuple $x^*$ forms an instance of $\text{OR}(\tilde{Q})$, and by the definition of $\text{OR}(\tilde{Q})$ we know that $x^* \in \text{OR}(\tilde{Q})$ if and only if some element of the tuple is contained in $\tilde{Q}$. By tracing back the series of equivalences we therefore find that $x^* \in \text{OR}(\tilde{Q})$ if and only if some input $x_i$ is a YES-instance of SAT. Since we can construct $x^*$ in polynomial time and $|x^*|$ is polynomial in $m$, we have constructed a weak distillation of SAT into $\text{OR}(\tilde{Q})$. By Theorem 6 this implies NP $\subseteq$ coNP/poly and proves the theorem. ◀

▶ **Corollary 10.** *If some set $L$ is NP-hard under Karp reductions and $L$ cross-composes into the parameterized problem $Q$ then there is no polynomial kernel for $Q$ unless $NP \subseteq coNP/poly$.*

A simple extension of Theorem 9 shows that cross-compositions also exclude the possibility of compression into a small instance of a different parameterized problem, a notion sometimes referred to as bikernelization [20, 21]. If an NP-hard set cross-composes into a parameterized problem $Q$, then unless NP $\subseteq$ coNP/poly there is no polynomial-time algorithm that maps an instance $(x, k)$ of $Q$ to an equivalent instance $(x', k')$ of *any* parameterized problem $P$ with $|x'|, k' \leq k^{O(1)}$.

## **4**    **Results Based on Cross-Composition**

In this section we apply the cross-composition technique to give kernelization lower bounds. We consider the problems FEEDBACK VERTEX SET, CHROMATIC NUMBER and CLIQUE under various parameterizations. The first parameter we consider is the vertex cover number of a graph $G$, i.e. the cardinality of a smallest set of vertices $Z \subseteq V(G)$ such that all edges of $G$ have at least one endpoint in $Z$. We show that CLIQUE, CHROMATIC NUMBER and WEIGHTED FEEDBACK VERTEX SET do *not* admit polynomial kernels parameterized by the size of a vertex cover unless NP $\subseteq$ coNP/poly.

We could also define the vertex cover number as the minimum number of vertex deletions needed to reduce a graph to an edgeless graph; hence the vertex cover number measures how far a graph is from being edgeless. Following the initiative of Cai [9] we may similarly define the deletion distance of a graph $G$ to a (co-)cluster graph as the minimum number of vertices that have to be deleted from $G$ to turn it into a (co-)cluster graph. Since (co-)cluster graphs have a very restricted structure, one would expect that a parameterization by (co-) cluster deletion distance leads to fixed-parameter tractability; indeed this is the case for many problems, since graphs of bounded (co-)cluster deletion distance also have bounded cliquewidth (Lemma 19). For the FEEDBACK VERTEX SET problem, which admits a polynomial kernel parameterized by the target size and hence by the vertex cover number, we show that the parameterizations by cluster deletion or co-cluster deletion distance do not admit polynomial kernels.

In Table 2 we give the known results for our subject problems with respect to the standard parameterization, which refers to the solution size. Since the problems we study are very well-known, we do not give a full definition for each one. Instead we give an educative example of how the parameter is reflected in an instance.

> CHROMATIC NUMBER parameterized by the size of a vertex cover
> **Instance:** A graph $G$, a vertex cover $Z \subseteq V(G)$, and a positive integer $\ell$.
> **Parameter:** The size $k := |Z|$ of the vertex cover.
> **Question:** Is $\chi(G) \leq \ell$, i.e., can $G$ be colored with at most $\ell$ colors?

For technical reasons we supply a vertex cover in the input of the problem, to ensure that well-formed instances can be recognized in polynomial time. The parameter to the problem claims a bound on the vertex cover number of the graph, and using the set $Z$ we may verify this bound. For FEEDBACK VERTEX SET parameterized by deletion distance to cluster graphs or co-cluster graphs, we also supply the deletion set in the input. These versions of the problem are certainly no harder to kernelize than the versions where a deletion set or vertex cover is not given.

| Problem name | Parameter | Param. complexity | Kernel size |
|---|---|---|---|
| Clique | clique | W[1]-hard        [14] | W[1]-hard        [14] |
| Feedback Vertex Set | feedback vertex set | FPT        [10] | $4k^2$ vertices        [24] |
| Chromatic Number | chromatic number | NP-h for $k \in O(1)$ | NP-h for $k \in O(1)$ |

**Table 2** Parameterized complexity and kernel size for some of the problems considered in this paper, with respect to the standard parameterization (i.e., target size).

## 4.1 Clique parameterized by Vertex Cover

An instance of the NP-complete Clique problem [18, GT19] is a tuple $(G, \ell)$ and asks whether the graph $G$ contains a clique on $\ell$ vertices. We use this problem for our first kernelization lower bound.

▶ **Theorem 11.** Clique *parameterized by the size of a vertex cover does not admit a polynomial kernel unless* $NP \subseteq coNP/poly$.

**Proof.** We prove the theorem by showing that Clique cross-composes into Clique parameterized by vertex cover; by Corollary 10 this is sufficient to establish the claim. We define a polynomial equivalence relation $\mathcal{R}$ such that all bitstrings which do not encode a valid instance of Clique are equivalent, and two well-formed instances $(G_1, \ell_1)$ and $(G_2, \ell_2)$ are equivalent if and only if they satisfy $|V(G_1)| = |V(G_2)|$ and $\ell_1 = \ell_2$. From this definition it follows that any set of well-formed instances on at most $n$ vertices each is partitioned into $O(n^2)$ equivalence classes. Since all malformed instances are in one class, this proves that $\mathcal{R}$ is indeed a polynomial equivalence relation.

We now give a cross-composition algorithm which composes $t$ input instances $x_1, \ldots, x_t$ which are equivalent under $\mathcal{R}$ into a single instance of Clique parameterized by vertex cover. If the input instances are malformed or the size of the clique that is asked for exceeds the number of vertices in the graph, then we may output a single constant-size NO instance; hence in the remainder we may assume that all inputs are well-formed and encode structures $(G_1, \ell), \ldots, (G_t, \ell)$ such that $|V(G_i)| = n$ for all $i \in [t]$ and all instances agree on the value of $\ell$, which is at most $n$. We construct a single instance $(G', Z', \ell', k')$ of Clique parameterized by vertex cover, which consists of a graph $G'$ with vertex cover $Z' \subseteq V(G')$ of size $k'$ and an integer $\ell'$.

Let the vertices in each $G_i$ be numbered arbitrarily from 1 to $n$. We construct the graph $G'$ as follows (see also Figure 1):

1. Create $\ell n$ vertices $v_{i,j}$ with $i \in [\ell]$ and $j \in [n]$. Connect two vertices $v_{i,j}$ and $v_{i',j'}$ if $i \neq i'$ and $j \neq j'$. Let $C$ denote the set of these vertices. It is crucial that any clique in $G'$ can only contain one vertex $v_{i,\cdot}$ or $v_{\cdot,j}$ for each choice of $i \in [\ell]$ respectively $j \in [n]$. Thus any clique contains at most $\ell$ vertices from $C$.

2. For each pair $1 \leq p < q \leq n$ of distinct vertices from $[n]$ (i.e., vertices of graphs $G_i$), create three vertices: $w_{p,q}$, $w_{p,\hat{q}}$, and $w_{\hat{p},q}$ and make them adjacent to $C$ as follows:

   a. $w_{p,q}$ is adjacent to all vertices from $C$,
   b. $w_{p,\hat{q}}$ is adjacent to all vertices from $C$ except for $v_{\cdot,j}$ with $j = q$, and
   c. $w_{\hat{p},q}$ is adjacent to all vertices from $C$ except for $v_{\cdot,j}$ with $j = p$.

   Furthermore we add all edges between vertices $w_{\cdot,\cdot}$ that correspond to distinct pairs from $[n]$. Let $D$ denote these $3\binom{n}{2}$ vertices. Any clique can contain at most one $w_{\cdot,\cdot}$ vertex for each pair from $[n]$.

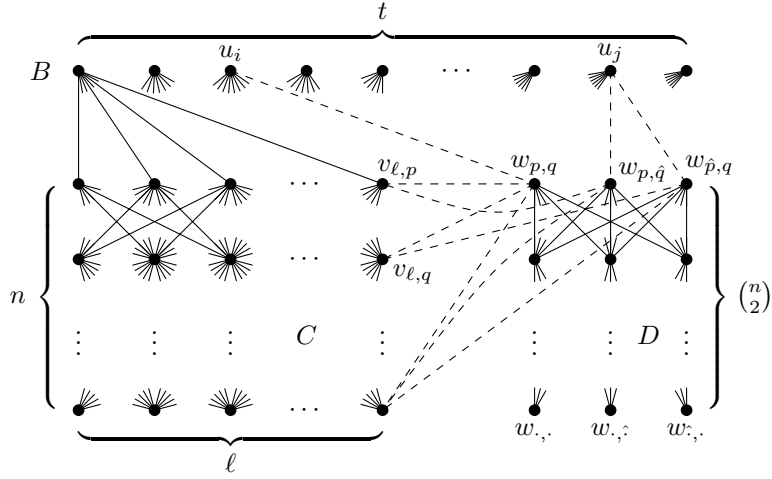

**Figure 1** A sketch of the construction used in the proof of Theorem 11. The dashed edges show in an examplary way how vertices $w_{p,q}$, $w_{p,\hat{q}}$, and $w_{\hat{p},q}$ are connected to vertices of $B$ and $C$, e.g., $\{p,q\}$ is an edge of $G_i$ but not of $G_j$.

3. For each instance $x_i$ with graph $G_i$ make a new vertex $u_i$ and connect it to all vertices in $C$. The adjacency to $D$ is as follows:

   a. Make $u_i$ adjacent to $w_{p,q}$ if $\{p,q\}$ is an edge in $G_i$.
   b. Otherwise make $u_i$ adjacent to $w_{p,\hat{q}}$ and $w_{\hat{p},q}$.

   Let $B$ denote this set of $t$ vertices.

We define $\ell' := \ell + 1 + \binom{n}{2}$. Furthermore, we let $Z' := C \cup D$ which is easily verified to be a vertex cover for $G'$ of size $k' := |Z'| = \ell n + 3\binom{n}{2}$. The value $k'$ is the parameter to the problem, which is polynomial in $n$ and hence in the size of the largest input instance. The cross-composition outputs the instance $x' := (G', Z', \ell', k')$. It is easy to see that our construction of $G'$ can be performed in polynomial time. Let us now argue that $x'$ is YES if and only if at least one of the instances $x_i$ is YES.

($\Leftarrow$) First we will assume that some $x_{i^*}$ is YES, i.e., that $G_{i^*}$ contains a clique on at least $\ell$ vertices. Let $S \subseteq [n]$ denote a clique of size exactly $\ell$ in $G_{i^*}$. We will construct a set $S'$ of size $\ell' = \ell + 1 + \binom{n}{2}$ and show that it is a clique in $G'$:

1. We add the vertex $u_{i^*}$ to $S'$.
2. Let $S = \{p_1, \ldots, p_\ell\} \subseteq [n]$. For each $p_j$ in $S$ we add the vertex $v_{j,p_j}$ to $S'$. By Step 1 all these vertices are pairwise adjacent, and by Step 3 they are adjacent to $u_{i^*}$.
3. For each pair $1 \le p < q \le n$ there are two cases:

   a. If $\{p,q\}$ is an edge of $G_{i^*}$ then the vertex $u_{i^*}$ is adjacent to $w_{p,q}$ in $G'$ (by Step 3) and $w_{p,q}$ is adjacent to all vertices of $C$ (by Step 2). We add $w_{p,q}$ to $S'$.
   b. Otherwise the vertex $u_{i^*}$ is adjacent to both $w_{p,\hat{q}}$ and $w_{\hat{p},q}$. Since the clique $S$ cannot contain both $p$ and $q$ when $\{p,q\}$ is a non-edge we are able to add $w_{p,\hat{q}}$ respectively $w_{\hat{p},q}$ to $S'$; recall that, e.g., $w_{p,\hat{q}}$ is adjacent to all vertices of $C$ except those corresponding to $q$.

   In both cases we add one $w_{\cdot,\cdot}$-vertex to $S'$, each corresponding to a different pair $p,q$; all these vertices are pairwise adjacent by Step 2.

We have identified the clique $S'$ in $G'$ of size $\ell' = \ell + 1 + \binom{n}{2}$, proving that $x'$ is a YES-instance.

($\Rightarrow$) Now assume that $x'$ is a YES-instance and let $S'$ be a clique of size $\ell + 1 + \binom{n}{2}$ in $G'$. Since $S'$ contains at most $\ell$ vertices from $C$ (i.e., one $v_{i,\cdot}$ for each $i \in [\ell]$) and at most $\binom{n}{2}$ vertices from $D$ it must contain at least one vertex from $B$, say $u_{i^*} \in B$. Since $B$ is an independent set the set $S'$ must contain exactly $\ell$ vertices from $C$ and exactly $\binom{n}{2}$ vertices from $D$. Let $S = \{j \in [n] \mid v_{i,j} \in S' \text{ for some } i \in [\ell]\}$. The set $S$ has size $\ell$ since $S'$ contains at most one vertex $v_{\cdot,j}$ for each $j \in [n]$. We will now argue that $S$ is a clique in $G_{i^*}$. Let $p, q \in S$. The clique $S'$ must contain a $w_{\cdot,\cdot}$-vertex corresponding to $\{p, q\}$ and it must contain vertices $v_{i,p}$ and $v_{i',q}$ for some $i, i' \in [\ell]$. Therefore it must contain $w_{p,q}$ since $w_{p,\hat{q}}$ has no edges to vertices $v_{\cdot,q}$ and $w_{\hat{p},q}$ has no edges to $v_{\cdot,p}$ by Step 2. Thus $u_{i^*} \in S'$ must be adjacent to $w_{p,q}$ which implies that $G_{i^*}$ contains the edge $\{p, q\}$. Thus $S$ is a clique in $G_{i^*}$.

Since we proved that the instance $(G', Z', \ell', k')$ can be constructed in polynomial-time and that it acts as the OR of the input instances, and because the parameter value $k'$ is bounded by a polynomial in the size of the largest input instance, this concludes the cross-composition proof and establishes the claim.                                         ◄

▶ **Corollary 12.** *If $\mathcal{F}$ is a class of graphs containing all cliques, then* VERTEX COVER *and* INDEPENDENT SET *parameterized by the minimum number of vertex deletions to obtain a graph in $\mathcal{F}$ do not admit polynomial kernels unless* $NP \subseteq coNP/poly$. *In particular,* VERTEX COVER *and* INDEPENDENT SET *parameterized by co-cluster deletion distance or cluster deletion distance do not admit polynomial kernels unless* $NP \subseteq coNP/poly$.                    ◄

## 4.2 Chromatic Number parameterized by Vertex Cover

In this section we give a kernelization lower bound for CHROMATIC NUMBER parameterized by vertex cover, through the use of a restricted version of 3-COLORING.

▶ **Definition 13.** A graph $G$ is a *triangle split graph* if $V(G)$ can be partitioned into sets $X, Y$ such that $G[X]$ is an edgeless graph and $G[Y]$ is a disjoint union of vertex-disjoint triangles.

An instance of the classical problem 3-COLORING WITH TRIANGLE SPLIT DECOMPOSITION is a tuple $(G, X, Y)$ consisting of a graph $G$ and a partition of its vertex set into $X \cup Y$ such that $G[X]$ is edgeless and $G[Y]$ is a union of vertex-disjoint triangles. The question is whether $G$ has a proper 3-coloring. The following lemma shows that this restricted form of the problem is NP-complete, which is proven by replacing all edges in a normal instance of 3-COLORING with a triangle. The proof is deferred to the appendix due to space restrictions.

▶ **Lemma 14.** 3-COLORING WITH TRIANGLE SPLIT DECOMPOSITION *is NP-complete.*    ◄

▶ **Theorem 15.** CHROMATIC NUMBER *parameterized by the size of a vertex cover does not admit a polynomial kernel unless* $NP \subseteq coNP/poly$.

**Proof.** To prove the theorem we will show that 3-COLORING WITH TRIANGLE SPLIT DECOMPOSITION cross-composes into CHROMATIC NUMBER parameterized by a vertex cover of the graph. By a suitable choice of polynomial equivalence relation in the same style as in Theorem 11 we may assume that we are given $t$ input instances which encode structures $(G_1, X_1, Y_1), \ldots, (G_t, X_t, Y_t)$ of 3-COLORING WITH TRIANGLE SPLIT DECOMPOSITION with $|X_i| = n$ and $|Y_i| = 3m$ for all $i \in [t]$ (i.e., $m$ is the number of triangles in each instance). We will compose these instances into one instance $(G', Z', \ell', k')$ of CHROMATIC NUMBER parameterized by vertex cover. By duplicating some instances we may assume that the number of inputs $t$ is a power of 2; this only increases the input size by a factor of at most 2, and hence any bounds which are polynomial in the old input size will be polynomial in the new input size which is sufficient for our purposes.

For each set $Y_i$, label the triangles in $G_i[Y_i]$ as $T_1, \dots, T_m$ in some arbitrary way, and label the vertices in each triangle $T_j$ for a set $Y_i$ as $a_i^j, b_i^j, c_i^j$. We build a graph $G'$ with a vertex cover of size $k' := 3 \log t + 4 + 3m \in O(m + \log t)$ such that $G'$ can be $\ell' := \log t + 4$-colored if and only if one of the input instances can be 3-colored.

1. Create a clique on vertices $\{p_i \mid i \in [\log t]\} \cup \{w, x, y, z\}$; it is called the *palette*.
2. Add the vertices $\bigcup_{i=1}^{t} X_i$ to the graph, and make them adjacent to the vertex $w$.
3. For $i \in [m]$ add a triangle $T_i^*$ to the graph on vertices $\{a_i, b_i, c_i\}$. The union of these triangles will be the *triangle vertices* $T^*$. Make all vertices in $T^*$ adjacent to all vertices from the set $\{p_i \mid i \in [\log t]\} \cup \{w\}$.
4. For $i \in [\log t]$ add a path on two new vertices $\{q_0^i, q_1^i\}$ to the graph, and make them adjacent to all vertices $(\{p_j \mid j \in [\log t]\} \cup \{x, y, z\}) \setminus \{p_i\}$. These vertices form the *instance selector* vertices.
5. For each instance number $i \in [t]$ consider the binary representation of the value $i$, which can be expressed in $\log t$ bits. Consider each position $j \in [\log t]$ of this binary representation, where position 1 is most significant and $\log t$ is least significant. If bit number $j$ of the representation of $i$ is a 0 (resp. a 1) then make vertex $q_0^j$ (resp. $q_1^j$) adjacent to all vertices of $X_i$. (We identify $t$ by the all-zero string $0 \dots 0$.)
6. As the final step we re-encode the adjacencies between vertices in the independent sets $X_i$ and the triangles into our graph $G'$. For each $i \in [t]$, for each vertex $v \in Y_i$, do the following. If $v$ is adjacent in $G_i$ to vertex $a_i^j$ then make vertex $v$ adjacent in $G'$ to $a_j$. Do the same for adjacencies of $v$ to $b_i^j$ and $c_i^j$.

This concludes the construction. The following claims about $G'$ are easy to verify:

(I) In every proper $\ell' = \log t + 4$-coloring of $G'$, the following must hold:

    a. each of the $\log t + 4$ vertices of the palette clique receives a unique color,

    b. consider some $i \in [\log t]$: the vertices $q_0^i$ and $q_1^i$ receive different colors (since they are adjacent), one of them must take the color of $w$ and the other of $p_i$ (they are adjacent to all other vertices of the palette),

    c. the triangle vertices $T^*$ are colored using the colors of $x, y, z$ (they are adjacent to all other vertices of the palette),

    d. the only colors which can occur on a vertex in $X_i$ (for all $i \in [t]$) are the colors given to $x, y, z$ and $\{p_j \mid j \in [\log t]\}$ (since the vertices in $X_i$ are adjacent to $w$).

(II) For every $i \in [t]$, the graph $G'[X_i \cup T^*]$ is isomorphic to $G_i$.

(III) The set $Z' := \{p_i \mid i \in [\log t]\} \cup \{w, x, y, z\} \cup T^* \cup \{q_0^i, q_1^i \mid i \in [\log t]\}$ forms a vertex cover of $G'$ of size $k' = |Z'| = 3 \log t + 4 + 3m$. Hence we establish that $G'$ has a vertex cover of size $O(m + \log t)$.

Due to space restrictions we cannot give the full correctness proof for the transformation. Using the given properties of $G'$ one may verify that $\chi(G') \leq \log t + 4 \Leftrightarrow \exists i \in [t] : \chi(G_i) \leq 3$. The full proof is in the appendix as Lemma 22. ◄

For every fixed integer $q$, the $q$-COLORING problem parameterized by the vertex cover number *does* admit a polynomial kernel. Kernelization algorithms for structural parameterizations of the $q$-COLORING problem will be the topic of a future publication.

### 4.3 Kernelization lower bounds for Feedback Vertex Set

In this section we give several kernelization lower bounds for FEEDBACK VERTEX SET. Due to space constraints the proofs are deferred to the appendix.

▶ **Theorem 16.** FEEDBACK VERTEX SET *parameterized by deletion distance to co-cluster graphs does not admit a polynomial kernel unless NP ⊆ coNP/poly.* ◀

▶ **Theorem 17.** FEEDBACK VERTEX SET *parameterized by deletion distance to cluster graphs does not admit a polynomial kernel unless NP ⊆ coNP/poly.* ◀

▶ **Theorem 18.** WEIGHTED FEEDBACK VERTEX SET, *where each vertex is given a positive integer as its weight, does not admit a polynomial kernel parameterized by the size of a vertex cover unless NP ⊆ coNP/poly.* ◀

## 5 Conclusions

We have introduced the technique of cross-composition and used it to derive kernelization lower bounds for structural parameterizations of several graph problems. Since we expect that cross-composition will be a fruitful tool in the further study of kernelization lower bounds, we give some pointers on how to devise cross-composition constructions. As the source problem of the composition one may choose a restricted yet NP-hard version of the target problem; this brings down the richness of the instances that need to be composed. If the goal is to give a lower bound for a structural parameterization (such as the size of a vertex cover) then starting from a problem on graphs which decompose into an independent set and some very structured remainder (e.g. triangle split graphs decompose into an independent set and vertex-disjoint triangles) it may be possible to compose the instances by taking the disjoint union of the inputs, and one-by-one identifying the vertices in the structured remainder. The fact that cross-compositions allow the output parameter to be polynomial in the size of the largest input can also be exploited, e.g., the proof of Theorem 11 uses this when composing input instances on $n$ vertices into a graph $G'$: we create $n^{O(1)}$ vertices inside a vertex cover $Z'$ for $G'$, and the adjacencies between $Z'$ and a single vertex outside the cover represent the entire adjacency structure of an input graph.

Cross-composition is also appealing from a methodological point of view, since it gives a unified way of interpreting the two earlier techniques for proving kernelization lower bounds: OR-compositions and polynomial-parameter transformations can both be seen to yield cross-compositions for a problem. For OR-composition this is trivial to see since an OR-composition for problem $Q$ just shows that the unparameterized variant $\tilde{Q}$ cross-composes into $Q$. The combination of an OR-composition for problem $P$ and a polynomial-parameter transform from $P$ to $Q$ also gives a cross-composition: first applying the OR-composition on instances of $P$ and then transforming the resulting $P$-instance to a $Q$-instance effectively shows that we can cross-compose instances of the unparameterized variant $\tilde{P}$ into instances of $Q$. Hence the cross-composition technique puts the existing methods of showing super-polynomial kernelization lower bounds in a common framework, and also explains *why* these problems do not admit polynomial kernels: a parameterized problem $P$ does not admit a polynomial kernel if it can encode the OR of *some* NP-hard problem for a sufficiently small parameter value. This new perspective might lead to a deeper insight into the common structure of FPT problems without polynomial kernels.

**Acknowledgements** We would like to thank Holger Dell for insightful discussions which led to a more elegant proof of Theorem 9.

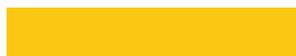

## A      Parameterized complexity of cluster and co-cluster deletion parameters

In this section we briefly show that FEEDBACK VERTEX SET is in FPT parameterized by cluster deletion or co-cluster deletion distance, through an argument about cliquewidth [22]. The following proposition about cliquewidth is folklore.

▶ **Proposition 1.** A cluster graph has clique-width 1.

We also use two results from Allen, Lozin and Rao [1].

▶ **Proposition 2.** For any graph $G$ it holds that $\text{CLIQUEWIDTH}(\overline{G}) \leq 2\,\text{CLIQUEWIDTH}(G)$, where $\overline{G}$ is the edge-complement of $G$.

▶ **Proposition 3.** If a graph $G$ is obtained from a graph $H$ by deleting $k$ vertices, then $\text{CLIQUEWIDTH}(G) \leq \text{CLIQUEWIDTH}(H) \leq 2^k(\text{CLIQUEWIDTH}(G) + 1)$.

These propositions allow us to relate the parameter "deletion distance to a (co-)cluster graph" to the cliquewidth of a graph.

▶ **Lemma 19.** *If graph $H$ can be turned into a cluster graph or co-cluster graph by $k$ vertex deletions, then $\text{CLIQUEWIDTH}(G) \leq 3 \cdot 2^k$.*

**Proof.** By Proposition 1 and Proposition 2 the cliquewidth of cluster graphs is 1, and the clique-width of co-cluster graphs is at most two. Assume $H$ can be turned into a cluster graph or co-cluster graph $G$ by exactly $k$ vertex deletions. Then $\text{CLIQUEWIDTH}(G) \leq 2$ and from Proposition 3 it follows that $\text{CLIQUEWIDTH}(H) \leq 2^k(\text{CLIQUEWIDTH}(G) + 1) \leq 2^k \cdot 3$.      ◀

Lemma 19 shows that graphs of bounded cluster graph deletion number or co-cluster graph deletion number, also have bounded cliquewidth. Since the FEEDBACK VERTEX SET problem can be solved in FPT-time on graphs of bounded cliquewidth [7], this shows that FEEDBACK VERTEX SET is in FPT when parameterized by cluster deletion distance or co-cluster deletion distance.

## B      Omitted proofs

### B.1      Proof of Corollary 12

**Proof of Corollary 12.** Consider an instance $(G, Z, \ell, k)$ of CLIQUE parameterized by the size of a vertex cover. Since a clique in $G$ is an independent set in $\overline{G}$, the CLIQUE instance is equivalent to asking whether the graph $\overline{G}$ has an independent set of size at least $\ell$. Because $Z$ is a vertex cover for $G$ we know that $G - Z$ is an independent set, and therefore $\overline{G} - Z$ is a clique. Hence if we use a parameter "deletion distance from a complete graph" which measures how many vertex deletions are needed to obtain a complete graph, then the instance $(G, Z, \ell, k)$ of CLIQUE parameterized by vertex cover is equivalent to an instance $(\overline{G}, Z, \ell, k)$ of INDEPENDENT SET parameterized by the size of the set $Z$ whose deletion from $\overline{G}$ leaves a complete graph. Since $\overline{G}$ has an independent set of size $\ell$ if and only if it has a vertex cover of size $|V(G)| - \ell$ it follows that these two instances are also equivalent to the instance $(\overline{G}, Z, |V(G)| - \ell, k)$ of VERTEX COVER parameterized by a deletion set $Z$ to a complete graph.

Since the proof of Theorem 11 shows that instances of CLIQUE cross-compose into an instance $(G, Z, \ell, k)$ of CLIQUE parameterized by vertex cover, and since this instance is equivalent to instance $(\bar{G}, Z, \ell, k)$ of INDEPENDENT SET parameterized by deletion distance to complete graphs and instance $(\overline{G}, Z, |V(G)| - \ell, k)$ of VERTEX COVER parameterized by deletion distance to complete graphs, this proves that Clique cross-composes into the latter two parameterized problems and hence they do not admit polynomial kernels unless NP $\subseteq$ coNP/poly.

Let $\mathcal{F}$ be a class of graphs containing all complete graphs. Then the minimum number of vertex deletions needed to transform a graph $G$ into a graph in $\mathcal{F}$ is at most the number of vertex deletions needed to turn $G$ into a complete graph. Hence the parameter "deletion distance to a graph in $\mathcal{F}$" is not larger than the parameter "deletion distance to complete graphs", and therefore Independent Set and Vertex Cover do not have a polynomial kernel for the parameter deletion distance to $\mathcal{F}$. Since the classes of cluster graphs and co-cluster graphs contain all cliques, this proves all claims in the corollary.                    ◀

## B.2    Proofs for Chromatic Number parameterized by Vertex Cover

An odd cycle is a simple cycle on an odd number of $\geq 3$ vertices. An odd wheel is the graph which is obtained from an odd cycle by adding a new vertex which is adjacent to all other vertices. The vertices on the odd cycle become the *rim* of the wheel, whereas the new universal vertex is the *hub* of the wheel. The following proposition about coloring odd wheels can be found in any standard text book on graph theory.

▶ **Proposition 4.** An odd wheel is not 3-colorable.

▶ **Lemma 20.** *Let $G$ be a graph and let $u$ and $v$ be distinct non-adjacent vertices in $G$ such that $G[N_G(\{u, v\})]$ contains an odd cycle. Then $u$ and $v$ must receive different colors in a proper 3-coloring of $G$.*

**Proof.** Proof by contradiction. Assume there is a proper 3-coloring of $G$ where $u$ and $v$ receive the same color. The coloring is still proper if we identify the vertices $u$ and $v$ into a single vertex $z$ which takes the same color as $u$ and $v$ (discarding parallel edges that might arise). After the transformation this new vertex $z$ is adjacent to all vertices in $N_G(\{u, v\})$. Since we assumed $G[N_G(\{u, v\})]$ contains an odd cycle, all vertices of this odd cycle are adjacent to $z$ after merging $u$ and $v$. But this shows that in the transformed graph $z$ forms the hub of an odd wheel with the vertices on the odd cycle as the rim. By Proposition 4 a graph containing an odd wheel cannot be 3-colored, which is a contradiction to the 3-coloring we extracted from the assumption that $G$ is 3-colored with the same color for $u$ and $v$; this proves the claim.                    ◀

▶ **Lemma 21.** 3-Coloring with Triangle Split Decomposition *is NP-complete.*

**Proof.** It is well-known that 3-Coloring on general graphs is NP-complete [18, GT4], and it is trivial to see that the problem restricted to triangle split graphs is contained in NP. We show how to transform an instance $G$ of 3-coloring in polynomial time into an equivalent instance of 3-coloring on a graph $G'$ with a triangle split decomposition of $V(G')$ into sets $X', Y'$. Number the edges in $G$ as $e_1, e_2, \ldots, e_m$. Construct the graph $G'$ as follows:
- Set $V(G') := V(G) \cup \{a_i, b_i, c_i \mid i \in [m]\}$.
- Add the edges $\{a_i, b_i\}, \{b_i, c_i\}, \{a_i, c_i\}$ to $E(G')$ for $i \in [m]$.
- For each edge $e_i = \{u_i, v_i\}$ ($i \in [m]$) of graph $G$, make vertex $u_i$ adjacent in $G'$ to $a_i$, and make $v_i$ adjacent to $b_i$ and $c_i$.
- Define $X' := V(G)$ and $Y' := \{a_i, b_i, c_i \mid i \in [m]\}$.

This concludes the description of $G'$. It is easy to see that $G'$ is a triangle split graph with the partition $X'$ and $Y'$ since $G'[X']$ is an independent set and $G'[Y']$ is a disjoint union of triangles. We now show that $\chi(G') \leq 3$ if and only if $\chi(G) \leq 3$.

($\Rightarrow$) Assume that $\chi(G') \le 3$ and consider a 3-coloring of $G'$. For every edge $\{u_i, v_i\} \in E(G)$ we added a triangle on vertices $\{a_i, b_i, c_i\}$ to the graph $G'$. Hence $G'[N_{G'}(\{u_i, v_i\})]$ contains an odd cycle for all pairs of vertices $\{u_i, v_i\}$ which are adjacent in $G$. By Lemma 20 this implies that $u_i$ and $v_i$ receive different colors in a 3-coloring of $G'$, and therefore the 3-coloring of $G'$ restricted to the vertex set of $G$ is a proper 3-coloring of $G$.

($\Leftarrow$) Assume that $G$ has a proper 3-coloring. We construct a 3-coloring for $G'$ by coloring all vertices of $V(G') \cap V(G)$ the same as in $G$; now all that remains is to color the triangles we added to the graph. If there is a triangle $\{a_i, b_i, c_i\}$ for a pair $\{u_i, v_i\}$ then $\{u_i, v_i\}$ are adjacent in $G$ and hence they receive different colors in the proper coloring. Now give $a_i$ the color of $v_i$, give $b_i$ the color of $u_i$ and give $c_i$ the remaining color. If we do this for every triangle then we obtain a proper 3-coloring of $G'$ which proves that $\chi(G') \le 3$.

Since the instance $(G', X', Y')$ can be built from $G$ in polynomial time this proves that 3-Coloring with Triangle Split Decomposition is NP-complete. ◀

▶ **Lemma 22.** *Let* $(G_1, X_1, Y_1), \ldots, (G_t, X_t, Y_t)$ *be input instances of* 3-Coloring with Triangle Split Decomposition *which are mapped to the instance* $(G', Z', \ell')$ *of* Chromatic Number *parameterized by vertex cover according to the construction of Theorem 15. Then* $\chi(G') \le \ell' \Leftrightarrow \exists i \in [t] : \chi(G_i) \le 3$.

**Proof.** Throughout the proof we will refer to the structural claims made about the graph $G'$ in the proof of Theorem 15.

($\Rightarrow$) Suppose $\chi(G') \le \ell'$ and consider some proper $\ell'$-coloring of $G'$. By (Ib) we know that for each $i \in [\log t]$ exactly one vertex of the pair $\{q_0^i, q_1^i\}$ receives the same color as $p_i$. Consider the string of $\log t$ bits where the $i$-th most significant bit is a 1 if and only if vertex $q_1^i$ receives the same color as $p_i$. This bitstring encodes some integer $i^* \in [t]$. We focus on the instance with the number $i^*$. Let $Q$ be the set of vertices which contains for each pair $\{q_0^i, q_1^i\}$ ($i \in [\log t]$) the unique vertex which is colored the same as $p_i$. By the definition of $G'$ we know that all vertices of $X_{i^*}$ are adjacent to all vertices of $Q$; hence in any proper coloring of $G'$ the vertices of $X_{i^*}$ cannot use any colors which are used on $\{p_i \mid i \in [\log t]\}$. By (Id) this implies that the coloring for $G'$ can only use the colors of $x, y, z$ on the vertices of $X_{i^*}$. By (Ic) the triangle vertices $T^*$ are also colored using only the colors of $x, y, z$. The graph $G'[X_{i^*} \cup T^*]$ is isomorphic to the input graph $G_{i^*}$ by (II), and since the coloring of $G'$ only uses the colors of $x, y, z$ on these vertices, this shows that the coloring of $G'$ restricted to the induced subgraph $G'[X_{i^*} \cup T^*]$ is in fact a 3-coloring of graph $G_{i^*}$, which proves that $\chi(G_{i^*}) \le 3$ and establishes this direction of the equivalence.

($\Leftarrow$) Suppose $\chi(G_{i^*}) \le 3$ for some $i^* \in [t]$. We will construct a proper $\ell'$-coloring of $G'$. Start by giving all vertices of the palette different colors. By (II) the graph $G'[X_{i^*} \cup T^*]$ is isomorphic to $G_{i^*}$. Re-label the colors in the 3-coloring of $G_{i^*}$ such that it uses the colors given to $\{x, y, z\}$ in our partial $\ell'$-coloring of $G'$. Give a vertex $v$ in the induced subgraph $G'[X_{i^*} \cup T^*]$ the same color as the vertex in $G_{i^*}$ to which it is mapped by the isomorphism. Afterwards we have a proper partial $\ell'$-coloring, where all vertices of the palette, all vertices of $X_{i^*}$, and all triangle vertices of $G'$ are colored. It remains to color the sets $X_i$ for $i \ne i^*$, and the pairs $\{q_0^i, q_1^i\}$. For each $i \in [\log t]$ we color the pair $\{q_0^i, q_1^i\}$ as follows: if the $i$-th most significant bit of the binary representation of the number $i^*$ is a 1 then we color $q_1^i$ the same color as $p_i$ and we color $q_0^i$ as $w$; if the bit is a 0 then we do it the other way around. It is straight forward to verify that we do not create any monochromatic edges in this way. As the final step we have to color the sets $X_i$ for $i \ne i^*$; so consider some $i \in [t]$ with $i \ne i^*$. The binary representation of the number $i^*$ must differ from the binary representation of $i$ in at least one position; suppose they differ at position $j$. The vertex of $\{q_0^j, q_1^j\}$ which matches the bit value of $i^*$ at position $j$ was colored the same as $p_j$, hence the other vertex of the pair must have been colored the same as $w$. Since the bit values differ, by the definition of adjacencies in $G'$ we find that the vertices $X_i$ are adjacent to the vertex of $\{q_0^j, q_1^j\}$ which is colored as $w$. Therefore the vertices of $X_i$ do not have any neighbors colored as $p_j$, and since $X_i$ is an independent set we may color all vertices in it the same as $p_j$. If we color all sets $X_i$ for $i \neq i^*$ in this way we obtain a proper $\ell'$-coloring of $G'$ which proves that $\chi(G') \leq \ell'$. ◄

## B.3 Feedback Vertex Set parameterized by Co-cluster Deletion Distance

In the proof of this section we will use the Feedback Vertex Set problem restricted to bipartite input graphs of girth at least six. An instance of the problem Feedback Vertex Set on Bipartite Graphs of Girth $\geq 6$ (FVS-BG6) is a tuple $(G, X, Y, \ell)$ and consists of a bipartite graph $G$ of girth at least 6 with bipartition of the vertex set into $X \cup Y$, and a target value $\ell$ and asks whether $G$ has a feedback vertex set of size at most $\ell$.

▶ **Observation 1.** Feedback Vertex Set on Bipartite Graphs of Girth $\geq 6$ is NP-complete. This follows from the fact that a normal instance of Feedback Vertex Set can be reduced to an equivalent instance on a bipartite graph of girth at least six by subdividing each edge with three new degree-2 vertices.

**Proof of Theorem 16.** We prove the theorem by showing that Feedback Vertex Set on Bipartite Graphs of Girth $\geq 6$ cross-composes into Feedback Vertex Set parameterized by deletion distance to co-cluster graphs (FVS-DCC); by Observation 1 and Corollary 10 this is sufficient to establish the claim. We start by defining a polynomial equivalence relation for our input instances. Using a standard encoding (such as an adjacency-list) it is easy to verify the bipartition and the bound on the girth in polynomial time. Hence we can test in polynomial time whether an instance is well-formed. We define our polynomial equivalence relation $\mathcal{R}$ such that all malformed instances are equivalent, and two well-formed instances $(G_1, X_1, Y_1, \ell_1)$ and $(G_2, X_2, Y_2, \ell_2)$ are equivalent if and only if they satisfy $|X_1| = |X_2|$, $|Y_1| = |Y_2|$ and $\ell_1 = \ell_2$. From this definition it follows that any set of well-formed instances on at most $n$ vertices each, is partitioned into $O(n^3)$ equivalence classes. Since all malformed instances are in one class, this proves that $\mathcal{R}$ is indeed a polynomial equivalence relation.

We now give a cross-composition algorithm which composes $t$ input instances $x_1, \ldots, x_t$ which are equivalent under $\mathcal{R}$ into a single instance of FVS-DCC. If the input instances are malformed then we may output a single constant-size no instance of FVS-DCC; hence in the remainder we may assume that all inputs are well-formed and encode structures $(G_1, X_1, Y_1, \ell), \ldots, (G_t, X_t, Y_t, \ell)$ which all agree on the value of $\ell$ and for which $|X_i| = |X_j|$ and $|Y_i| = |Y_j|$ for all $i, j \in [t]$. We now construct in polynomial time a single instance $(G', Z', \ell', k')$ of FVS-DCC which is yes if and only if one of the input instances is yes, and such that $k' = |Z'|$ is bounded by $|Y_1|$; since the maximum size of an input instance is at least $|Y_1|$ this will show that the parameter size satisfies the requirements for a cross-composition.

We construct the graph $G'$ starting from a disjoint union of the graphs $G_i$. We label the vertices in each set $Y_i$ arbitrarily from 1 to $n$, and then identify the vertex sets $Y_1, \ldots, Y_t$ to a new vertex set $Y^*$: we identify the first vertex of each set into one new vertex, the second vertex of each set, etc. We add all edges between vertex sets $X_i, X_j$ for all $i \neq j$. We observe that $G'[X_1 \cup \cdots \cup X_t]$ is a co-cluster graph. Thus $G'$ has a deletion distance of at most $|Y^*| = |Y_1|$ to co-cluster graphs. Let $\ell' := (t-1)|X_1| + \ell$, let the deletion set to co-cluster graphs be $Z' := Y^*$ which implies that the parameter to this problem is $k' := |Z'| = |Y^*|$. From this construction it follows that $G_i$ is isomorphic to $G'[X_i \cup Y^*]$. It remains to prove correctness of the cross-composition: the output instance $(G', Z', \ell', k')$ is YES if and only if one of the input instances is YES.

($\Leftarrow$) Let us first assume that some instance, say $x_i$, is YES and let $S$ be a feedback vertex set of $G_i$ of size at most $\ell$. Let $S' := S \cup \bigcup_{j \neq i} X_j$. It is easy to see that $G' - S' = G_i - S$ and that $|S'| \leq (t-1)|X_1| + \ell$. Thus the output instance is YES.

($\Rightarrow$) Let us now assume that the output instance is YES and let $S'$ be a feedback vertex set for $G'$ of size at most $\ell' = (t-1)|X_1| + \ell$. We first observe that $S'$ must completely contain almost all sets $X_i$. Indeed, if there are three sets $X_{i_1}, X_{i_2}, X_{i_3} \nsubseteq S'$ then $G' - S'$ contains a triangle since we added all edges between different sets $X_i$, $X_j$.

If there is exactly one set $X_i$ with $X_i \nsubseteq S'$ then $S'$ contains $\bigcup_{j \neq i} X_j$. Letting $S := S' \setminus \bigcup_{j \neq i} X_j$ we observe that $G' - S' = (G' - \bigcup_{j \neq i} X_j) - S = G_i - S$. Thus $S$ is a feedback vertex set of $G_i$, since $G_i - S = G' - S'$ is acyclic by choice of $S'$. Furthermore, since $\bigcup_{j \neq i} X_j \subseteq S'$ we get that $|S| \leq |S'| - (t-1)|X_1| \leq \ell$. Thus $x_i$ is a YES-instance of FVS-BG6.

It remains to consider the case that there are two sets $X_i, X_j$ with $X_i, X_j \nsubseteq S'$. If $S'$ misses at least two vertices in each of the two sets then $G' - S'$ would contain a cycle of length four. Thus we assume w.l.o.g. that $|X_j \setminus S'| = 1$ and we let $u$ denote the vertex of $X_j$ that is not in $S'$. We recall that $u$ is adjacent to all vertices of $X_i$. If $S'$ does not contain any vertex from $X_i \cup Y^*$ then $G_i = G'[X_i \cup Y^*]$ is acyclic. In that case $x_i$ is a YES-instance of FVS-BG6 and we are done. Otherwise let $v \in S' \cap (X_i \cup Y^*)$ and let $S'' := (S' \setminus \{v\}) \cup \{u\}$. We will show that $S''$ is a feedback vertex set of $G'$ (of size at most $\ell'$) and with $\bigcup_{j \neq i} X_j \subseteq S''$ which permits us to reuse the argument from the previous paragraph to show that $x_i$ is YES.

We assume for contradiction that $G' - S''$ is not acyclic. Thus there must be a cycle $C$ which contains the vertex $v$. Since $\bigcup_{j \neq i} X_j \subseteq S''$ the cycle $C$ is contained in a copy of $G_i$ implying that it has length of at least six. We let $C = (\dots, p, q, v, r, s, \dots)$ and consider two cases:

- $v \in Y^*$: Since $G_i$ is bipartite, the vertices $q$ and $r$ must be in $X_i$ and are adjacent to $u \in X_j$ by construction. Thus $C' = (\dots, p, q, u, r, s, \dots)$ would be a cycle in $G' - S'$. A contradiction.

- $v \in X_i$: In this case $p$ and $s$ must be in $X_i$ and adjacent to $u \in X_j$, implying that $C' = (\dots, p, u, s, \dots)$ would be a cycle in $G' - S'$. A contradiction.

Thus $S''$ is a feedback vertex set for $G'$ of size at most $\ell'$ and with $\bigcup_{j \neq i} X_j \subseteq S''$. By the previous argumentation this implies that $x_i$ is a YES instance.

Since it is easy to verify that the instance $(G', Z', \ell', k')$ can be constructed in polynomial time from the input instances, this establishes all components required for the cross-composition and concludes the proof. ◀

## B.4 Feedback Vertex Set parameterized by Cluster Deletion Distance

▶ **Definition 23.** The $K_4$-in-a-box graph $B_{K_4}$ (see Figure 2) is the graph obtained from a complete graph on 4 vertices $\{a, b, c, d\}$ by adding a new degree-2 vertex $v$ for each pair $\{a, b\}, \{b, c\}, \{c, d\}, \{d, a\}$ such that $v$ is adjacent to both vertices of the pair. The vertices $\{a, c\}$ are the 0-labeled terminals of the graph, and the vertices $\{b, d\}$ are the 1-labeled terminals of the graph.

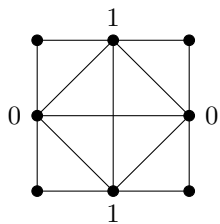

■ **Figure 2** The $K_4$-in-a-box graph $B_{K_4}$ with labeled vertices.

It is straight-forward to verify that any feedback vertex set for $B_{K_4}$ has size at least 2, and that a size-2 feedback vertex set contains either the 0-labeled terminals or the 1-labeled terminals.

**Proof of Theorem 17.** We give a cross-composition from INDEPENDENT SET to FEEDBACK VERTEX SET parameterized by deletion distance from cluster graphs. Let $x_1, \ldots, x_t$ be $t$ instances of INDEPENDENT SET each coming with a graph $G_i$ on $n$ vertices and $m$ edges and asking for an independent set of size at least $\ell$. W.l.o.g. we assume $t$ to be a power of two. We also consider the vertices of the graphs $G_i$ to be numbered arbitrarily from 1 to $n$. For the cross-composed instance we construct a graph $G'$ as follows:

1. We add an independent set on $n$ vertices, labeled $v_1, \ldots, v_n$. Let $B$ denote this independent set. It is intended to encode the selection of a feedback vertex set.

2. To build an instance selector we use $\log t$ copies of the $B_{K_4}$ graph. Each copy corresponds to one of the $\log t$ bit positions necessary to express numbers from 1 to $t$ (by convention $t$ corresponds to $0 \ldots 0$). The idea is to encode instance selection by forcing either the two 0-labeled terminals or the two 1-labeled terminals into the feedback vertex set. We make a total of $n$ copies of this construction.

3. For each instance $x_i$ and any edge $\{p, q\}$ of $G_i$ we make the following construction which is intended to check edges of the selected instance:

   a. Add a clique on $\log t + 2$ new vertices $w_1, \ldots, w_{\log t}, w_{\text{out}}, w_{\text{in}}$. (The vertex $w_{\text{out}}$ will be adjacent to vertices outside of the clique, namely in the independent set; vertex $w_{\text{in}}$ will only have neighbors inside the edge checker.)

   b. For each bit position $j \in [\log t]$, connect the vertex $w_j$ to the two vertices labeled 0 in the $j$-th $B_{K_4}$ graph of each instance selector if the $j$-th bit of $i$ is zero and to those with label 1 otherwise.

   c. Connect $w_{\text{out}}$ to $p$ and $q$ in $B$.

   We add $n + 2$ disjoint copies of this construction for each of the $m$ edges of each of the $t$ graphs.

We define $\ell' := 2n \log t + (n+2)tm \log t + (n - \ell)$ and we let $Z'$ contain $B$ as well as the $8n \log t$ vertices of the instance selectors. Clearly $G' - Z'$ is a disjoint union of cliques (namely the edge checkers), and the size of $Z'$ is $k' := n + 8n \log t$. The cross-composition creates the instance $x' = (G', Z', \ell', k')$, i.e., it asks whether the graph $G'$ has a feedback vertex set of size at most $\ell'$, and provides a deletion set $Z'$ (of size $k'$) such that $G' - Z'$ is a disjoint union of cliques. Clearly the parameter value $k'$ is polynomial in $\max |x_i| + \log t$ fulfilling the definition of a cross-composition. It is also easy to see that the construction can be performed in polynomial time. We will now argue correctness of the cross-composition, i.e., we will show that $G'$ has a feedback vertex set of size at most $\ell'$ if and only if at least one of the graphs $G_i$ has an independent set of size at least $\ell$.

It is helpful to observe that any feedback vertex set of $G'$ contains at least $2 \log t$ vertices from each instance selector (each of the $\log t$ $B_{K_4}$ graphs has two disjoint cycles) and at least $\log t$ vertices from each edge checker (since each of them is a clique on $\log t + 2$ vertices). Thus the minimum size of any feedback vertex set for $G'$ is at least $2n \log t + (n+2)tm \log t$.

($\Leftarrow$) We begin by assuming that some instance, say $x_{i^*}$, is YES, i.e., that $G_{i^*}$ has an independent set of size at least $\ell$. Let $S$ be an independent set of size $\ell$ in $G_{i^*}$; we will construct a feedback vertex set $S'$ of $G'$ of size at most $\ell'$:

1. In the independent set $B$ we select for $S'$ all vertices except for the $\ell$ vertices from $S$, for a total of $n - \ell$ vertices.

2. In each $j$-th $B_{K_4}$ graph in any instance selector, select the two 0-labeled terminals if the $j$-th bit in the binary expansion of $i^*$ is one, and select the two 1-labeled terminals otherwise. Thus we pick $2 \log t$ vertices per selector, i.e., $2n \log t$ vertices in total; clearly deleting these vertices takes care of any cycles inside the instance selectors.

3. In each edge checker that does not belong to $x_{i^*}$, say it corresponds to some instance $x_i$, we pick all vertices except for $w_{\text{in}}$ and some vertex, say $w_j$, where the binary expansions of $i^*$ and $i$ differ. Thus $w_j$ will not have neighbors in the instance selectors in $G' - S'$. We pick a total of $(n+2)(t-1)m \log t$ vertices; skipping $w_{\text{in}}$ and some single $w_j$ in each of these edge checkers.

4. For the edge checkers that correspond to $x_{i^*}$ we select all vertices except $w_{\text{out}}$ and $w_{\text{in}}$. These two remaining vertices have no neighbors in the instance selectors. Furthermore, in $G' - S'$ the vertices $w_{\text{out}}$ have degree one since they are adjacent to the endpoints of some edge $\{p, q\}$ from $G_{i^*}$ but we picked all $n$ vertices except for those from the independent set $S$, which cannot cannot contain both $p$ and $q$. Thus we pick a total of $2tm$ vertices.

Thus the set $S'$ is a feedback vertex set of $G'$ of size $2n \log t + (n+2)tm \log t + (n - \ell) = \ell'$, proving that the cross-composed instance is YES too.

($\Rightarrow$) Let us now assume that the cross-composed instance is YES and let $S'$ be a feedback vertex set of size $\ell' = 2n \log t + (n+2)tm \log t + (n - \ell)$ for $G'$. If $\ell = 0$ then trivially all instances $x_i$ are YES and there would be nothing to show; so assume $\ell > 0$. Since $S'$ contains at least $2 \log t$ vertices from each instance selector and at least $\log t$ vertices from each edge checker, it can contain at most $n - \ell$ vertices of the independent set $B$. Let $S$ denote those vertices of $B$ that were not chosen by $S'$; clearly $|S| \geq \ell$ as $|B| = n$.

We first observe that $S'$ cannot select more than 2 vertices per graph from $s$ $B_{K_4}$ graphs in instance selectors for any $s \geq n$. Otherwise, using the lower bounds for $S'$ on instance selectors and edge checkers, this would imply that the size of $S'$ is greater than $2n \log t + (n+2)tm \log t + (n - \ell)$; a contradiction to the choice of $S'$:

$$2n \log t + (n+2)tm \log t + s > 2n \log t + (n+2)tm \log t + (n - \ell),$$

since $s \geq n > n - \ell$. By the same argumentation $S'$ cannot select more than $\log t$ vertices in $n$ or more edge checkers. Now, considering that $G'$ contains $n$ copies of the instance selector, there must be at least one copy where $S'$ selects exactly 2 vertices in each of the $\log t$ $B_{K_4}$ graphs. Let $i^* \in \{1, \ldots, t\}$ be the number whose inverted binary expansion matches that selection of $S'$ (i.e., if $S'$ contains the two 0-vertices then the $j$-th bit of $i^*$ must be one); again, by convention $t$ matches $0 \ldots 0$. We will show that $S$ constitutes an independent set for $G_{i^*}$.

We begin by showing that $S'$ does not contain the $w_{\text{out}}$-vertex of at least two edge checkers of any edge $\{p, q\}$ of $G_{i^*}$ (recall that $w_{\text{out}}$ is adjacent to $p, q \in B$ in $G'$). The reason is that each $w_j$-vertex is connected to two terminals of a $B_{K_4}$ graph in which $S'$ selected the other two terminals, by which $w_j$ is not disconnected. Therefore, $S'$ must contain also all $w_j$-vertices in edge checkers for the graph $G_{i^*}$, else there would be a cycle through a $B_{K_4}$ graph and the $w_j$-vertex in $G' - S'$. Hence, by the previous argument, there are two edge checkers for each edge of $G_{i^*}$ where $S'$ does not select $w_{\text{out}}$.

Let $p, q \in S$; we show that $G_{i^*}$ does not contain the edge $\{p, q\}$: Assume for contradiction that $\{p, q\}$ is an edge of $G_{i^*}$ and recall that for at least two of the corresponding edge checkers $S'$ does not contain $w_{\text{out}}$. Thus $S'$ must contain $p$ or $q$, since otherwise there would be a cycle through $p$, $q$, and the $w_{\text{out}}$-vertices of the two checkers. This is a contradiction since $S$ is defined to contain all vertices of $B$ that are not in $S'$, i.e., $S = B \setminus S'$. Hence $S$ is indeed an independent set of $G_{i^*}$. ◄

## B.5 Weighted Feedback Vertex Set parameterized by Vertex Cover

An instance of the Weighted Feedback Vertex Set parameterized by vertex cover problem is a tuple $(G, Z, \ell, w, k)$ where $G$ is a graph, $Z$ is a vertex cover of $G$, $k = |Z|$, $\ell$ is a positive integer and $w : V(G) \to \mathbb{N}^+$ is a weight function that assigns a positive integral weight to every vertex. The question is whether $G$ has a feedback vertex set $S$ such that $\sum_{v \in S} w(v) \leq \ell$.

**Proof of Theorem 18.** We prove the theorem by showing that Feedback Vertex Set on Bipartite Graphs cross-composes into Weighted Feedback Vertex Set parameterized by a vertex cover. By a suitable choice of polynomial equivalence relation we may assume the input consists of well-formed instances $(G_1, X_1, Y_1, \ell), \ldots, (G_t, X_t, Y_t, \ell)$ which all agree on the number of vertices in $|X|$ and $|Y|$ and which have the same target value $\ell$. By the argument given in the proof of Theorem 15 we may assume that $t$ is a power of 2.

In each instance $i \in [t]$ we number the vertices of $X_i$ in an arbitrary way from 1 to $|X_i|$, and we also number $Y_i$ from 1 to $|Y_i|$. We construct a graph $G'$ with weight function $w'$ which has a vertex cover $Z'$ of size $k' := 8r + |X_1|$, and which has a feedback vertex set of total weight $\ell' := r(2t|X_1|) + (t-1)|X_1| + \ell$ if and only if one of the input graphs has a feedback vertex set of size $\ell$. We will define the weight function $w'$ in an informal way, by describing the weights that various sets of vertices should receive.

1. Add all independent sets $X_i$ for $i \in [t]$ to the new graph $G'$, and give these vertices weight 1.
2. Add a vertex set $Y^* = \{y_1, \ldots, y_{|Y_1|}\}$ to the graph and give each vertex weight 1. For each set $X_i$ with $i \in [t]$ and vertex $v_p \in X_i$ which is numbered $p$, for each neighbor of $v_p$ in $G_i$ numbered $q$ add the edge $\{v_p, y_q\}$ to $G'$. Observe that afterwards the graph $G'[X_i \cup Y^*]$ is isomorphic to $G_i$ for all $i \in [t]$.
3. We can represent an instance number in the range $[t]$ using exactly $\log t$ bits since we assumed $t$ is a power of 2. For each bit position $j \in [\log t]$ we create a copy of the graph $B_{K_4}$ described in Definition 23. We label its 0-terminal vertices $\{b_{j,0'}, b_{j,0''}\}$ and the 1-terminal vertices $\{b_{j,1'}, b_{j,''}\}$. For each instance number $i$ whose $j$-th most significant bit in the binary expansion is a 0, we make all vertices of $X_i$ adjacent to the 0-terminals $\{b_{j,0'}, b_{j,0''}\}$, and for instance numbers whose bit value is 1 we make it adjacent to $\{b_{j,1'}, b_{j,1''}\}$. We set the weight of each vertex in each copy of $B_{K_4}$ to $t|X_1|$.

This concludes the description of the graph $G'$ and weight function $w'$. Since a valid instance of Weighted Feedback Vertex Set parameterized by vertex cover also contains a vertex cover set $Z'$, we must supply such a vertex cover as part of the output of the procedure. It is easy to verify that if we let $Z'$ contain the vertex set $Y^*$ and all vertices of each of the $\log t$ copies of $B_{K_4}$ then this forms a vertex cover of size $|X_1| + 8 \log t$, hence we can use this set as part of the output. The parameter value is the size of this vertex cover, and its value $k' := |Z'| = |X_1| + 8 \log t$ is bounded by a polynomial in $\log t$ plus the size of the largest instance. It is easy to see that this construction can be carried out in polynomial time. It remains to prove that $G'$ has a feedback vertex set of weight $\ell'$ if and only if one of the input graphs has a feedback vertex set of size $\ell$.

($\Rightarrow$) Assume that $G'$ has a feedback vertex set $S'$ of total weight at most $\log t(2t|X_1|) + (t-1)|X_1| + \ell$. The graph $G'$ contains $\log t$ vertex-disjoint copies of the graph $B_{K_4}$. By Definition 23 the feedback vertex set $S'$ must contain at least two vertices from each copy of $B_{K_4}$. If there is some copy of $B_{K_4}$ from which $S'$ contains more than two vertices, then this set must have weight at least $3t|X_1| + (\log t - 1)2t|X_1|$; but then the set $S''$ which contains all 0-terminal vertices of the copies of $B_{K_4}$ and the vertices $\bigcup_{i=1}^{t} X_i$ has weight $\log t(2t|X_1|) + t|X_1|$ which is at most as large. Hence by updating the set $S'$ we may assume that it contains exactly two vertices from each copy of $B_{K_4}$, and from Definition 23 it then follows that for each copy it contains either the 0-terminal vertices or the 1-terminal vertices. We now construct the binary representation of an instance number using the contents of $S'$. Let the $j$-th bit of the number be a 1 if set $S'$ contains $\{b_{j,1'}, b_{j,1''}\}$, and a 0 in the case that it contains $\{b_{j,0'}, b_{j,0''}\}$; let $i^*$ denote the instance number in the range $[t]$ which is represented by this bitstring. Observe that by the choice of $i^*$, for all vertices in $X_{i^*}$ all of their neighbors in the $B_{K_4}$ graphs are contained in $S'$. On the other hand, if we consider some instance number $i' \neq i^*$ then there is at least one bit position where the representations of the numbers $i'$ and $i^*$ differ. Let $j$ be such a bit position and assume for the moment that the $j$-th bit of the number $i^*$ is a 1, which implies the $j$-th bit of $i'$ is a 0 (the other case is symmetric). Then $S'$ contains the terminal vertices $\{b_{j,1'}, b_{j,1''}\}$ but does not contain $\{b_{j,0'}, b_{j,0''}\}$. But then $S'$ must contain all vertices from the set $X_{i'}$, for if $S'$ would avoid some vertex $v \in X_{i'}$ then the graph $G' - S'$ would contain a cycle on vertices $\{v, b_{j,0'}, b_{j,0''}\}$ which contradicts the assumption that $S'$ is a feedback vertex set. This shows that for all instance numbers $i' \neq i^*$ the set $S'$ must contain all vertices of $X_{i'}$. These vertices together with the 2 terminal vertices in each copy of $B_{K_4}$ account for $\log t(2t|X_1|) + (t-1)|X_1|$ of the weight of $S'$, and therefore the remaining vertices in $S'$ have weight at most $\ell$; in particular the set $S'$ contains at most $\ell$ vertices from the set $X_{i^*} \cup Y^*$ since each such vertex has weight 1. We observed earlier that the graph $G'[X_{i^*} \cup Y^*]$ is isomorphic to $G_{i^*}$. Since $S'$ is a feedback vertex set for $G'$ it must also break all cycles in all induced subgraphs, hence $G'[X_{i^*} \cup Y^*] - S'$ is acyclic. But since $S'$ contains at most $\ell$ vertices from $X_{i^*} \cup Y^*$ this proves that $S' \cap (X_{i^*} \cup Y^*)$ is a feedback vertex set of size at most $\ell$ for graph $G_{i^*}$.

($\Leftarrow$) Assume that $G_{i^*}$ has a feedback vertex set $S$ of size $\ell$ for some input graph $G_{i^*}$. We show how to construct a feedback vertex set $S'$ for $G'$ of weight at most $\ell' = \log t(2t|X_1|) + (t-1)|X_1| + \ell$.

1. For each bit position $j \in [\log t]$ add $\{b_{j,0'}, b_{j,0''}\}$ to $S'$ if the $j$-th bit of $i^*$ is a 0, and otherwise add $\{b_{j,1'}, b_{j,1''}\}$. This contributes a total weight of $r(2t|X_1|)$.
2. Add the set $\bigcup_{i \neq i^*} X_i$ for a total weight of $(t-1)|X_1|$.
3. Finally add the vertices from $X_{i^*}$ and $Y^*$ which correspond to the vertices in $S$; this adds a total weight of $|S| = \ell$.

Hence the resulting set $S'$ has weight exactly $\ell'$. To see that $S'$ is indeed a feedback vertex set for $G'$, observe that by taking two matching terminal vertices for each copy of $B_{K_4}$ we have broken all cycles within the $B_{K_4}$ graphs. For all sets $X_{i'}$ with $i' \neq i^*$ we have taken all the $X_{i'}$ vertices in $S'$ so $G' - S'$ cannot contain cycles through such sets $X_{i'}$. By taking the appropriate terminal vertices in $S'$ we have broken all connections between vertices in $X_{i^*}$ and vertices in copies of $B_{K_4}$. Finally there can be no cycles in $G'[X_{i^*} \cup Y^*] - S'$ since we assume that $S$ is a feedback vertex set for $G_{i^*}$ which is isomorphic to $G'[X_{i^*} \cup Y^*]$, and we have made the same choices as $S$ to break all cycles in that induced subgraph. Hence $S'$ is indeed a feedback vertex set of the desired weight.

We have proven that our newly constructed instance $(G', Z', \ell', w', k')$ indeed acts as the OR of instances $x_1, \ldots, x_t$. Since the output parameter $k := |C|$ is appropriately bounded, this shows the correctness of the cross-composition. By invoking Corollary 10 this is sufficient to show that WEIGHTED FEEDBACK VERTEX SET parameterized by a vertex cover does not admit a polynomial kernel unless NP $\subseteq$ coNP/poly.                                    ◀